\documentclass[useAMS,usenatbib]{mn2e}
\usepackage[dvips]{graphicx}
\usepackage{fixltx2e}
\usepackage{epstopdf}
\usepackage{times}
\usepackage{amsmath}
\usepackage{color}
\newcommand{\aap}{A\&A}

\newcommand{\mnras}{MNRAS}

\newcommand{\apj}{ApJ}

\newcommand{\aj}{AJ}

\title[Binary star evolution]{Evolution of Binary Stars in Multiple-Population Globular Clusters}
\author[J. Hong et al.]  {Jongsuk Hong$^1$\thanks{E-mail: hongjong@indiana.edu}, Enrico Vesperini$^1$, Antonio Sollima$^2$, Stephen L.W. McMillan$^3$,\\ \newauthor Franca D'Antona$^4$ and Annibale D'Ercole$^2$.\\
  $^1$Department of Astronomy, Indiana University, Bloomington, IN, 47401, USA\\
  $^2$INAF - Osservatorio Astronomico di Bologna, via Ranzani 1, I-40127 Bologna, Italy\\
  $^3$Department of Physics, Drexel University, Philadelphia, PA 19104, USA\\
  $^4$INAF - Osservatorio Astronomico di Roma, via di Frascati 33, I-00040 Monteporzio, Italy\\
}
\begin{document}

\newcommand{\srA}{{MPr5f03}}
\newcommand{\srB}{{MPr10f03}}
\newcommand{\srC}{{MPr5f1}}
\newcommand{\srD}{{SPf1}}
\newcommand{\xg}{x_{\rm g,0} }

\maketitle

\label{firstpage}

\begin{abstract}
The discovery of multiple stellar populations in globular clusters has implications for all the aspects of the study of these stellar systems. 
 In this paper, by means of $N$-body simulations, we study the
 evolution of binary stars in multiple-population clusters and explore
 the implications of the initial differences in the spatial
 distribution of different stellar populations for the evolution and
 survival of their binary stars. Our simulations show that initial
 differences between the spatial distribution of first-generation (FG)
 and second-generation (SG) stars can leave a fingerprint in the
 current properties of the binary  population. SG binaries are
 disrupted more efficiently than those of the FG population  resulting
 in a  global SG binary fraction smaller than that of the FG.   As
 for surviving binaries, dynamical evolution  produces a difference between
 the SG and the FG binary binding energy distribution with the SG
 population characterized by a larger fraction of high binding energy
 (more bound) binaries. We have also studied the dependence of the binary
 properties on  the distance from the cluster centre. Although the
 global binary fraction decreases more rapidly for the SG population, the local binary fraction measured in the cluster inner regions
 may still be dominated by SG binaries. The
 extent of the differences between the surviving  FG and SG binary
 binding energy distribution also varies radially within the cluster and
 is larger in the cluster inner regions.    
\end{abstract}


\section{Introduction}
\label{sec:intro}

Numerous observational studies have revealed that  globular
clusters host multiple stellar populations characterized by different
chemical (see e.g.  Carretta et al. 2009a, 2009b, Gratton et al. 2012 and references therein) and photometric properties
(see e.g. Lee et al. 1999, Bedin et al. 2004, Ferraro et al. 2004, Piotto et al. 2007, 2015; Milone et al. 2008, 2010, 2012; Bellini et al. 2013). The implications
of the presence of multiple populations in globular clusters and the
questions raised by this discovery span all the aspects of the
astrophysics of these stellar systems including their formation,
chemical and dynamical evolution along with the possible role played
by these objects in the assembly of the Galactic halo.

A complete scenario for the formation and evolution of multiple-population
globular clusters requires the identification of the source(s) of gas with
chemical properties consistent with the observed abundance patterns
and out of which second-generation (SG) stars might have formed along with
models following the dynamics of this gas, the SG formation
history and the subsequent early and long-term dynamical evolution of
the multiple-population cluster.
Different sources of gas for the SG formation have been suggested in the
literature (see e.g. Ventura et al. 2001, Prantzos \& Charbonnel 2006, Decressin et al. 2007, de Mink et al. 2009, Bastian et al. 2013) and a number of studies have addressed some of the issues
concerning the origin of the observed abundance patterns (see e.g. Decressin et al. 2007, D'Ercole et al. 2008, 2010, 2012), the
formation and dynamical evolution of multiple-population clusters (see e.g. D'Ercole et al. 2008, Bekki 2011, Vesperini et al. 2010, 2013, Decressin et al. 2008, 2010). 
The theoretical study of multiple-population clusters is still in
its infancy and most questions and challenges are just now beginning to be addressed.

In D'Ercole et al. (2008), we have modelled the formation and dynamical
evolution of the SG population forming from the ejecta of
first-generation (FG)
intermediate-mass AGB stars; one of the predictions of our study was
that the SG population would form in a
compact subsystem concentrated in the inner regions of the FG
cluster. In a subsequent study on the long-term evolution of
multiple-population clusters (Vesperini et al. 2013) we have shown
that some memory of the initial spatial segregation of the SG
population should be preserved in many Galactic clusters and, indeed, several
observational studies (see e.g. Sollima et al. 2007, Bellini et al. 2009, Lardo et al. 2011, Milone et al. 2012, Beccari et al. 2013, Johnson \& Pilachowski
2012, Cordero et al. 2014, Kucinskas et al. 2014) have found clusters in which SG stars are 
more spatially concentrated than FG stars (see also Bellazzini et al. 2012 and Richer et al. 2013 for  interesting studies on the kinematical properties of FG and SG stars).  

One of the dynamical implications of the SG spatial segregation
concerns the evolution of binary stars. Until SG and FG stars are
completely mixed, SG stars will evolve in a denser environment where
disruption and evolution of SG binaries will occur more rapidly than
for FG binaries. In Vesperini et al. (2011) we have carried out an
initial investigation of this issue by combining  $N$-body and
semi-analytical calculations of the disruption of binaries in
multiple-population clusters. Our study showed that SG binary
disruption rate can indeed be significantly larger than that of FG
binaries and  that the overall binary disruption rate in
multiple-population clusters is larger than that of a single
population cluster without the central SG dense subsystem. 

An early
investigation by D'Orazi et al. (2010), although based on a small
sample of Ba stars, suggests that the fraction of SG binaries might indeed be
smaller than that of FG binaries.

As for the connection with observational studies, as already pointed out in Vesperini et
al. (2011) and further discussed in more detail in this paper, it is
important to emphasize the differences  between the theoretical predictions concerning the global
binary fraction and the local binary fraction measured at different distances from
the cluster centre. Specifically, we will show in this paper that even if the global SG binary fraction
is smaller than the FG one, the local SG binary fraction measured in the
cluster innermost regions may still be larger than the FG one.

In this paper we extend the initial investigation by Vesperini et al. (2011)  by carrying out a
survey of $N$-body simulations and including in a self-consistent way a
population of binary stars instead of just relying on analytical estimates
for binary disruption rates due to single/binary
interactions. The $N$-body simulations presented here, while still
idealized in several aspects, allow us to follow in much more detail the evolution
of the cluster binary population and its dynamics. The questions we address in this
paper concern
the radial variation of the fraction of FG and SG binary stars
resulting from the combined effect of disruption and tendency of
binary stars to segregate in the cluster inner regions, the evolution
of the
global binary fraction and its relation with the local fraction
measured at different distances from the cluster centre, and
the evolution of the binding energy of the surviving binaries.

The results of our investigation show that differences in the initial structural
properties of the FG and SG populations can leave a significant fingerprint in the
properties of FG and SG binaries.

The outline of this paper is the following. In Section 2 we describe
the method and the initial conditions adopted in our simulations. In
section 3 we present our results, and in section 4 we summarize the main conclusions of our investigation.

\section{Methods and Initial Conditions}

The $N$-body simulations presented in this paper have been carried out
with the GPU-accelerated version of the code {\sc nbody6} (Aarseth 2003, Nitadori \& Aarseth 2012) and run on the {\sc big red ii} cluster at Indiana University.

In the initial conditions we have explored, the FG and SG subsystems have initially the same total mass and they both follow a King model (1966) density profile with central dimensionless potential
$W_0=7$; the SG subsystem, however, is initially concentrated in the inner regions of the FG system as suggested by the results of
D'Ercole et al. (2008). All the simulations start with equal-mass particles. We have explored two models here: one in which the initial
ratio of the FG to SG half-mass radius ($R_{\rm h,FG}/R_{\rm h,SG}$) is equal to 5, and one with $R_{\rm h,FG}/R_{\rm h,SG}=10$. In order to explore the differences between single-population and multiple-population clusters we have also run a number of simulations of single-population (SP) systems with the same number of particles and tidal radius as the multiple-population systems but with an internal structure described by a single King model with $W_0=7$.

We include the effects of the host galaxy tidal field (modelled as a point mass) and assume that the cluster is initially tidally truncated; particles moving beyond a radius equal to two
times the tidal radius are removed from the simulation.
\begin{table}
  \begin{center}
  \caption{Initial parameters for all models.}
  \begin{tabular}{l c c c c c c}
  \\
    \hline
    \hline
    Model id. & $N$ & $\frac{R_{\rm h,FG}}{R_{\rm h,SG}}$ & $f_{\rm b,0}$ & $x_{\rm g,0}$ \\
    \hline
    \srA{x3} & 20,000 & 5 & 0.03 & 3\\
    \srA{x5} & 20,000 & 5 & 0.03 & 5\\
    \srA{x10} & 20,000 & 5 & 0.03 & 10\\
    \srA{x20} & 20,000 & 5 & 0.03 & 20\\
    \srB{x3} & 20,000 & 10 & 0.03 & 3\\
    \srB{x5} & 20,000 & 10 & 0.03 & 5\\
    \srB{x10} & 20,000 & 10 & 0.03 & 10\\
    \srB{x20} & 20,000 & 10 & 0.03 & 20\\
    \srC{x3} & 20,000 & 5 & 0.1 & 3\\
    \srC{x5} & 20,000 & 5 & 0.1 & 5\\
    \srC{x10} & 20,000 & 5 & 0.1 & 10\\
    \srC{x20} & 20,000 & 5 & 0.1 & 20\\
    \srC{x3-20} & 20,000 & 5 & 0.1 & 3-20$^a$\\
    \hline
    \srD{x3} & 20,000 & - & 0.1 & 3\\
    \srD{x5} & 20,000 & - & 0.1 & 5\\
    \srD{x10} & 20,000 & - & 0.1 & 10\\
    \srD{x20} & 20,000 & - & 0.1 & 20\\
    \srD{x3-20} & 20,000 & - & 0.1 & 3-20$^a$\\
    \hline
    \srA{x10n32k} & 32,000 & 5 & 0.03 & 10\\
    \srA{x10n40k} & 40,000 & 5 & 0.03 & 10\\
    \srA{x10n60k} & 60,000 & 5 & 0.03 & 10\\
    \hline    
    \label{tbl1}
  \end{tabular}
  \end{center}
\begin{flushleft} 
$N=N_{\rm s}+N_{\rm bin}$ is the total number of single, $N_{\rm s}$, and binary, $N_{\rm bin}$, particles.\\
$f_{\rm b,0}=N_{\rm bin}/(N_{\rm s}+N_{\rm bin})$ is the initial binary fraction.\\
$\frac{R_{\rm h,FG}}{R_{\rm h,SG}}$ is the ratio of the FG to the SG initial  half-mass radii.\\
All the SP models refer to single-population systems.\\
$\xg$ initial hardness parameter (see section 2 for definition).\\
$^a$ A uniform distribution in binding energy is assumed.\\
\end{flushleft}
\end{table}

As far as the properties of the binary population are concerned we have explored the
evolution of systems with values of the initial binary fraction, $f_{\rm b,0}=N_{\rm bin}/(N_{\rm s}+N_{\rm bin})$ equal to 0.03
and 0.10 (where $N_{\rm s}$ is the number of single particles and $N_{\rm bin}$ is the number of binaries; we will refer to $N=N_{\rm bin}+N_{\rm s}$ to indicate the sum of the total number of single and binary particles; the total number of particles in each simulation is equal to $N_{\rm tot}=N_{\rm s}+2N_{\rm bin}$). 
All the simulations start with $N=20,000$ ($N_{\rm tot}=22,000$ for simulations with $f_{\rm b,0}=0.1$, and $N_{\rm tot}=20,600$ for simulations with $f_{\rm b,0}=0.03$). 
We assume that FG and SG binaries have the same initial binding energy distribution. In this paper we focus our attention on moderately hard binaries with  initial global hardness parameter, $\xg$, ranging from 3 to 20; $\xg$ is defined as $E_{\rm b}/(m \sigma_{\rm SP}^2)$ where $E_{\rm b}$ is the absolute value of the binary binding energy, and $\sigma_{\rm SP}$ the 1-D velocity dispersion of all stars in the SP system.
In order to better illustrate the dependence of the binary evolution on the hardness parameter, we have run a number of simulations each one including binaries with a single value of   $\xg$.
We have also carried out one simulation including binaries with a uniform binding energy distribution between $\xg=3$ and  $\xg=20$.

The initial parameters of all models and the ids. used to identify them in the rest of this paper are summarized in Table \ref{tbl1}. Some of the simulations performed reach a deep core collapse phase; specifically the following systems reach the deep core collapse phase
(MPr5f03x3,
MPr5f03x5,
MPr10f1x3,
MPr10f1x5,
MPr10f1x10,
MPr10f1x20,
MPr5f1x3,
MPr5f1x5,
SPf1x3)
at $t/\tau_{\rm rh,0}$ equal to about, respectively, 
(8.4,
8.7,
4.9,
4.9,
5.7,
7.8,
7.3,
8.3,
15.5). 

 In order to provide an indication of the extent of the stochastic variations in the results obtained we have carried out 10 simulations with different random realizations of the initial conditions for the system MPr5f03x5 and 10 for the system MPr5f03x20.
Three simulations with $N=32,000, 40,000$ and $60,000$  and structural properties equal to those of
MPr5f03x10 have also been carried out to explore the dependence of the results on the number
of particles.

\section{Results}

\subsection{Evolution of the global binary fraction}

\begin{figure}
  \includegraphics[width=85mm]{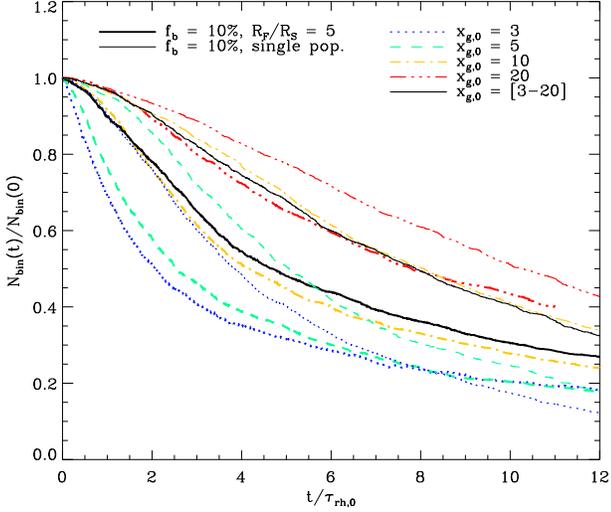}
  \caption{Time evolution of the total number of binaries, $N_{\rm bin}(t)$ (normalized to the total initial number of binaries, $N_{\rm bin}(0)$) for simulations MPr5f1 and SPf1 and different values of the hardness factor $\xg$. Time is normalized to the initial half-mass relaxation time, $\tau_{\rm rh,0}$.}
  \label{fig_tbin}
\end{figure}

Fig. \ref{fig_tbin} shows the time evolution of the global binary
fraction for the multiple-population systems MPr5f1x3, MPr5f1x5, MPr5f1x10, MPr5f1x20, MPr5f1x3-20, and the corresponding single-population systems SPf1x3, SPf1x5, SPf1x10, SPf1x20, SPf1x3-20.
As shown by this figure, multiple-population systems are characterized by a general enhancement in the binary disruption compared to single-population
clusters.
As discussed in Vesperini et al. (2011), the differences in the structural properties between
multiple-population and single-population clusters are responsible for
the observed differences in the evolution of the global binary fraction.

\begin{figure}
  \includegraphics[width=85mm]{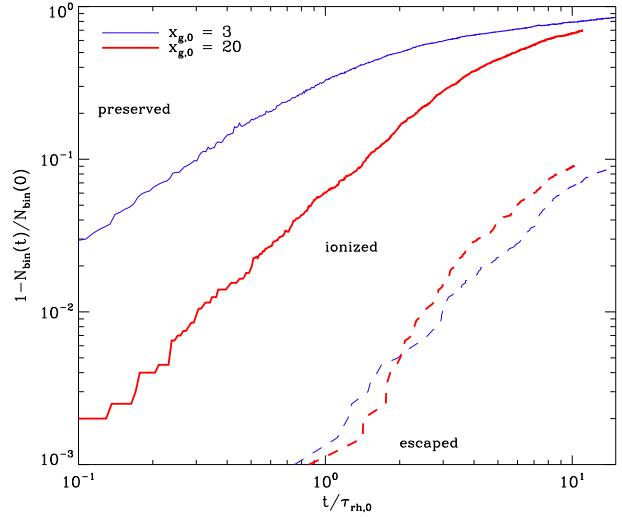}
  \caption{Time evolution (with time normalized to the initial half-mass relaxation time) of the total number of binaries (normalized
    to the total initial number of binaries). At any given time the
    difference between $1$ and the value shown by the solid line
    represents the fraction of the initial binary population
    surviving in the cluster, the difference between the  values
    shown by the solid line and the dashed line represents the
    fraction of the initial number of binaries ionized, and the dashed line shows the fraction of the initial number of binaries escaped from the cluster. Thick, red lines refer to
    the MPr5f1x20 simulation; thin blue lines refer to the MPr5f1x3
    simulation.} 
  \label{fig_esc}
\end{figure}

The role of different processes responsible for the decrease in the number of binaries is illustrated in Fig. \ref{fig_esc} for the simulations MPr5f1x3 and MPr5f1x20. For the range of binding energies considered in this paper, binary ionization is the dominant process while binary escape  play only a minor role.
Ionization is more efficient in the central high-density regions populated mainly by SG binaries and, as we will further discuss below, this leads to a significant preferential disruption of SG binaries. As expected, simulations with binaries with smaller values of $x_{\rm g,0}$ are characterized by a more rapid ionization of a larger fraction of binaries.

\begin{figure}
  \includegraphics[width=85mm]{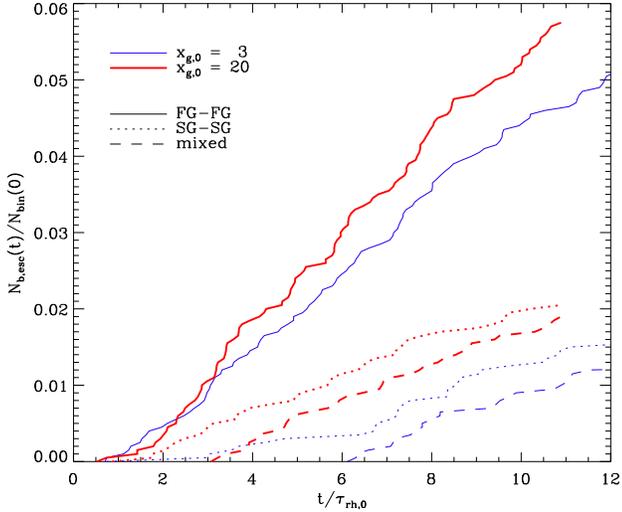}
  \caption{Time evolution (time is normalized to the initial half-mass relaxation time) of the total number of escaping binaries (normalized to the total
    initial number of binaries, $N_{\rm bin}(0)$).Thick, red lines refer to the MPr5f1x20
    simulation; thin blue lines refer to the MPr5f1x3
    simulation. Solid, dotted, and dashed lines refer, respectively,
    to FG, SG, and  mixed binaries.}
  \label{fig_kck}
\end{figure}

\begin{figure*}
  \includegraphics[width=170mm]{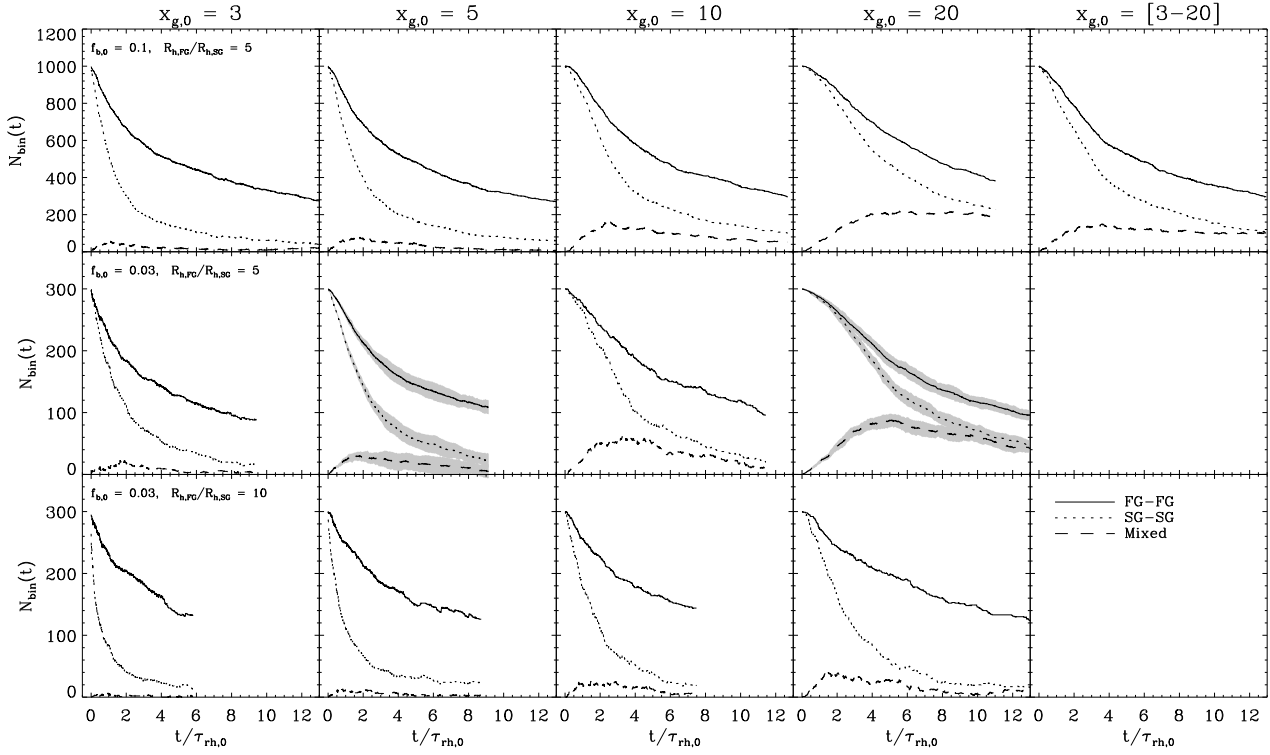}
  \caption{Time evolution of the total number of FG (solid lines), SG
    (dotted lines) and mixed (dashed lines)  binaries for all the
    MPr5f1 (first row), MPr5f03 (second row), and MPr10f03 (third row) simulations.
Each column shows the results of simulations for a different value of
$\xg$ indicated at the top of the figure. 
For simulations MPr5f03x5 and MPr5f03x20 the lines show the mean of the number of binaries from the results of the ten simulations with different random realizations of the initial conditions and the shaded area show the $\pm 1\sigma$. Time is normalized to the initial half-mass relaxation time. }
  \label{fig_bin}
\end{figure*}

\begin{figure}
  \includegraphics[width=85mm]{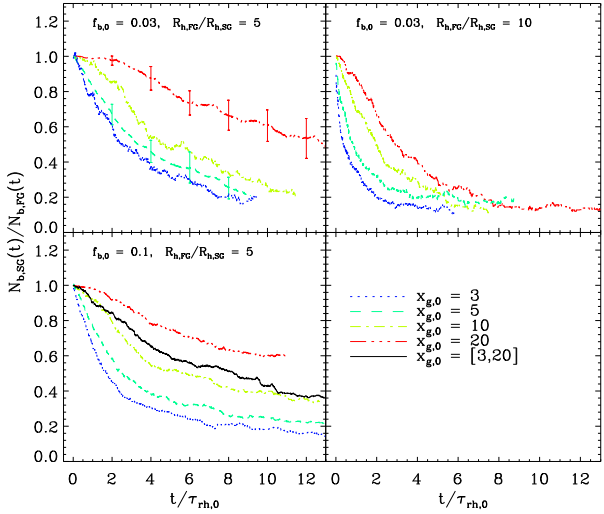}
  \caption{
  Time evolution (with time normalized to the initial half-mass relaxation time) of the SG-to-FG binary number ratio for MPr5f03 (upper left panel), MPr10f03 (upper right panel), and MPr5f1 (lower left panel) for different values of $\xg$ (see lower right panel for colors and line styles corresponding to the different values of $\xg$). For simulations MPr5f03x5 and MPr5f03x20 in the upper left panel, the lines show the mean of the SG-to-FG binary number ratio from the results of the 10 simulations with different random realizations of the initial conditions; bars showing the $\pm 1\sigma$ are also shown at a few representative times.
  }
  \label{fig_brt}
\end{figure}

\begin{figure}
  \includegraphics[width=85mm]{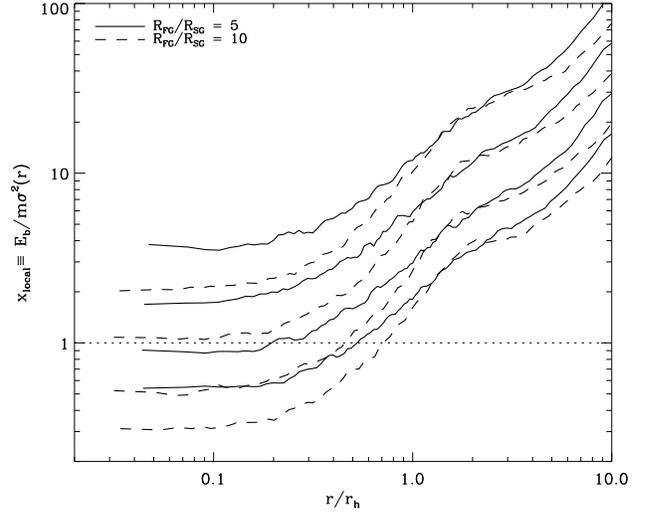}
  \caption{Radial profiles of initial local hardness parameter, $x_{\rm local}(r)$ (see \S 3.1 for definition) for the MPr5f03 (solid lines) and the MPr10f03 (dashed lines) simulations. Different  solid (and dashed) lines correspond (from top to bottom) to $\xg=$ 20, 10, 5, 3. Radius is normalized to the half-mass radius, $r_{\rm h}$.}
  \label{fig_hd}
\end{figure}

\begin{figure*}
  \includegraphics[width=175mm]{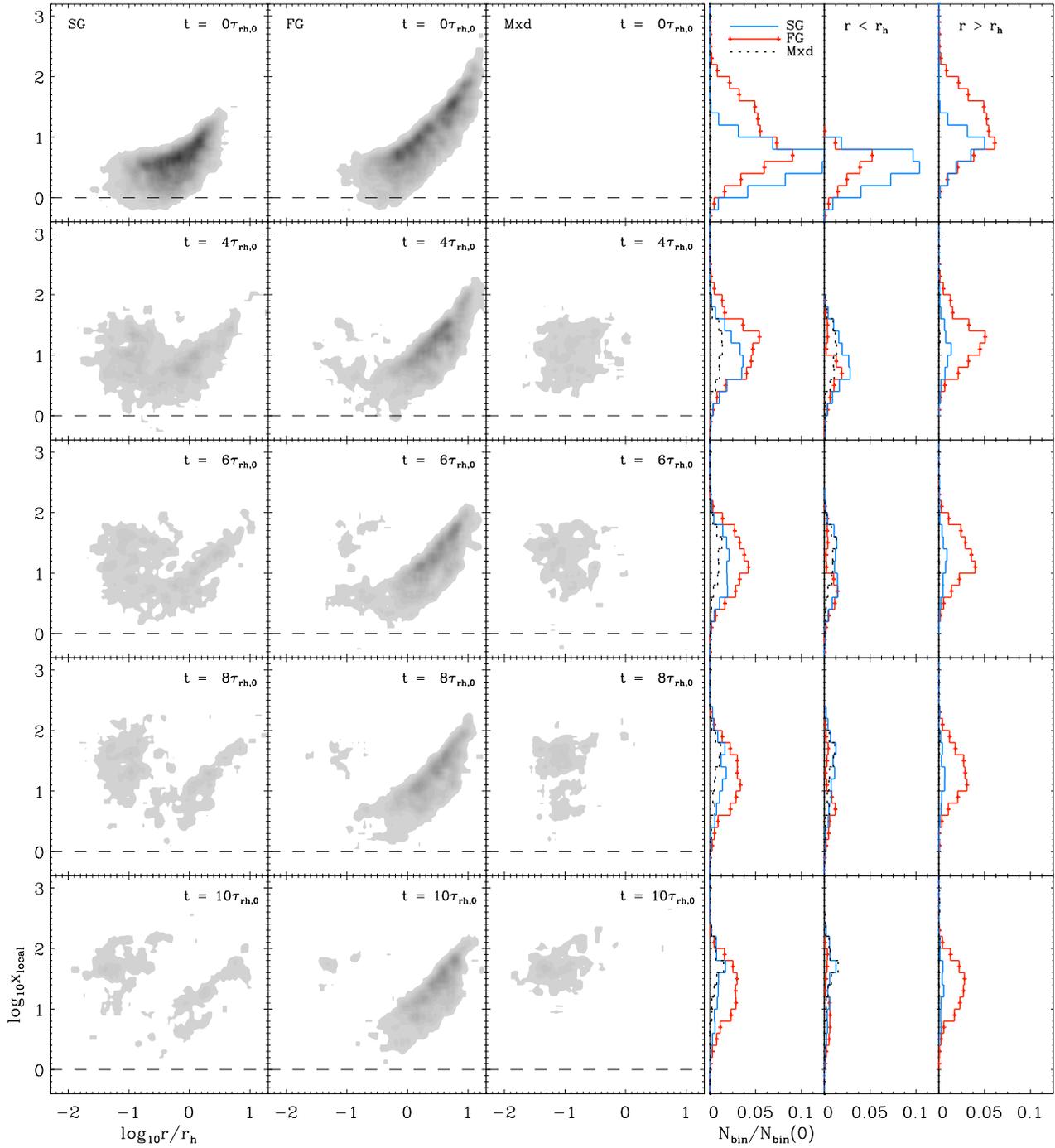}
  \caption{Time evolution of the density of SG (first column), FG (second column), and mixed (third column) binaries in the $\log(r/r_{\rm h})-\log(x_{\rm local}(r))$ plane (where $x_{\rm local}(r)$ is the local hardness parameter; see \S 3.1 for definition) for the MPr5f1x3-20 simulation. The last three columns of panels (on the right side of the figure) show the histograms of the distribution of $\log(x_{\rm local}(r))$ for SG (solid blue lines), FG (red lines and short vertical segments), and mixed binaries (dashed lines) for the whole system, for binaries within the half-mass radius and outside the half-mass radius. Data from three combined snapshots around the indicated time are used for these plots.
}
  \label{fig_xr2}
\end{figure*}

\begin{figure*}
  \includegraphics[width=175mm]{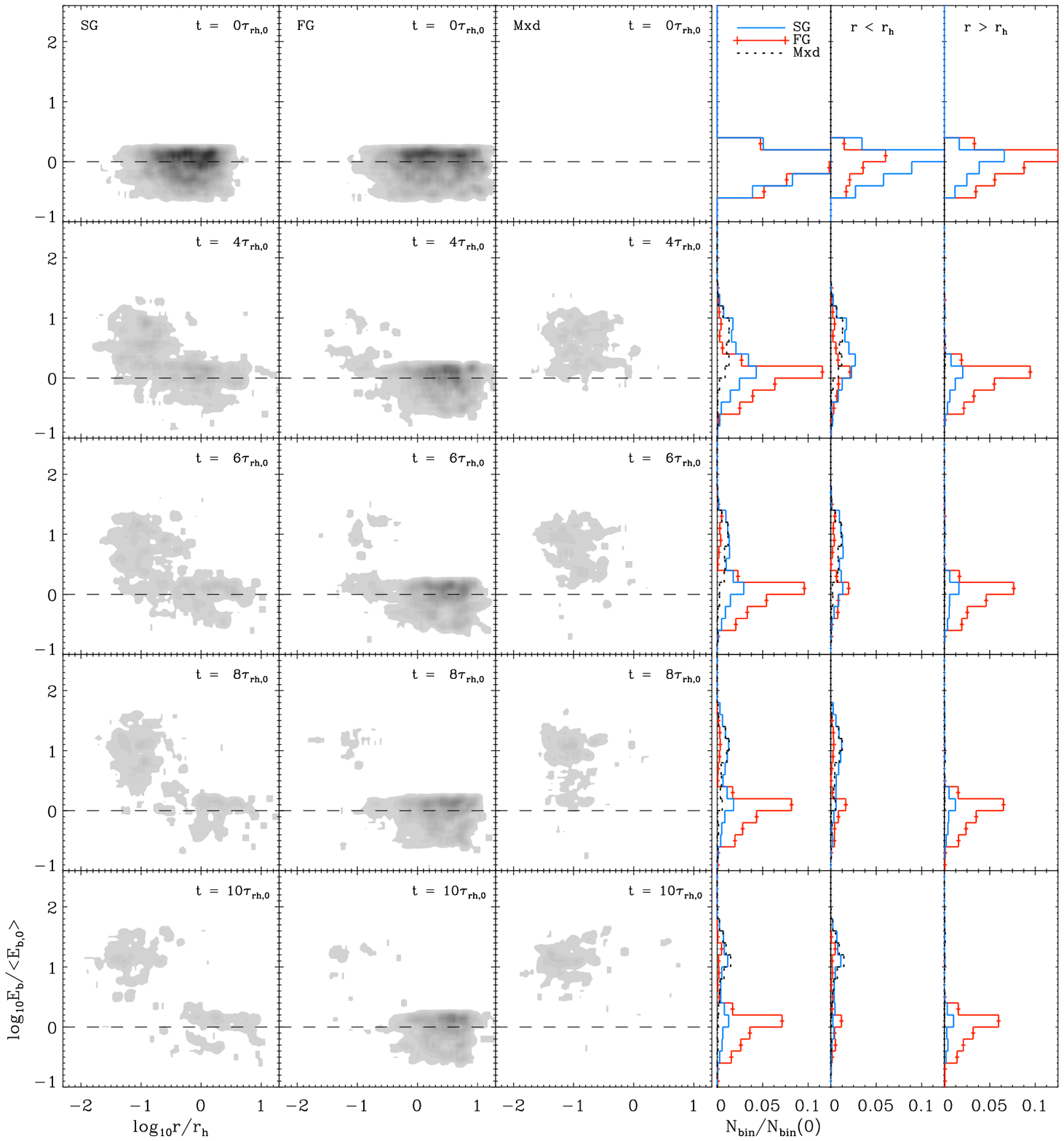}
  \caption{Time evolution of the density of SG (first column), FG (second column), and mixed (third column) binaries in the $\log(r/r_{\rm h})-\log(E_{\rm b}/\langle E_{\rm b,0}\rangle)$ plane (where $E_{\rm b}$ is the binary binding energy and $\langle E_{\rm b,0}\rangle$ is the initial mean binding energy of all binaries for the MPr5f1x3-20 simulation. The last three columns of panels (on the right-hand side of the figure) show the histograms of the distribution of $\log(E_{\rm b}/\langle E_{\rm b,0}\rangle)$ for SG (solid blue lines), FG (red lines and short vertical segments), and mixed binaries (dashed lines) for the whole system, for binaries within the half-mass radius and outside the half-mass radius. Data from three combined snapshots around the indicated time are used for these plots.}
  \label{fig_ebr2}
\end{figure*}

Binary escape affects primarily FG binaries but the preferential escape of FG binaries is far from being sufficient to balance the preferential ionization of SG binaries with the net result that the global fraction of SG binaries declines much more rapidly.
Fig.  \ref{fig_kck} show the number of FG, SG and mixed binaries (hereafter we refer to binaries that, as a result of
a component exchange event, have one SG and one FG component as mixed binaries) escaping for the MPr5f1x3 and the MPr5f1x20 simulations.
This figure further illustrates the role of escape in determining the decrease of the global binary fraction and shows the extent to which SG, FG and mixed binaries are affected by this process.

In Fig. \ref{fig_bin} we show the time evolution of total number of SG and FG and mixed
binaries. This figure clearly
illustrates the preferential disruption of SG binaries in all the
systems considered in this paper; it is also interesting to note that
exchange events, particularly for simulations with the harder
binaries, lead to the formation of a non-negligible number of mixed
binaries.

A comparison of the panels of Fig. \ref{fig_bin} showing the results for the MPr5  and the MPr10 simulations, show that, as was to be expected, systems starting with a more concentrated SG subsystem are characterized by a more efficient disruption of SG binaries. Even for the $\xg =20$ case, only about 6 \% of the initial population of SG binaries survive after about $10 \tau_{\rm rh,0}$ in the MPr10f03x20 simulation whereas in the less concentrated MPr5f03x20 system about 15 \% of the initial SG binary population survives after the same number of initial half-mass relaxation times.
Fig. \ref{fig_brt} shows the time evolution of the ratio of the total number of SG to the total number of FG binaries and further illustrates the preferential disruption of SG binaries.
For the simulations MPr5f03x5 and MPr5f03x20, the plots in Fig.  \ref{fig_bin} and \ref{fig_brt} show, respectively, the mean values of the number of binaries and the mean of the SG-to-FG binary number ratio along with the $1\sigma$ variation about the mean as calculated from ten different realizations of each of these two initial conditions.

In Fig. \ref{fig_hd} we show the radial profile of the local hardness parameter
$x_{\rm local}(r)$, defined as the ratio of the binary binding energy to 
$m\sigma(r)^2$where $\sigma(r)$ is the 1-D velocity dispersion measured at a distance $r$ from the cluster 
centre. This parameter provides a more specific measure of binary
hardness and of the expected dynamical fate of a binary as a function of the distance from the cluster centre. SG binaries
are concentrated in the cluster inner regions where the larger
velocity dispersion results in smaller values of $x_{\rm local}(r)$ and a more 
rapid disruption.

 For binaries in the cluster outer regions
(preferentially FG binaries), on the other hand, $x_{\rm local}$ is large and binaries can easily survive. FG binaries initially orbiting in the
cluster outer regions can be disrupted only later in the cluster
evolution after migrating towards the cluster central regions  as a result
of the mass segregation process. 

Fig.\ref{fig_xr2} and \ref{fig_ebr2} show the density of FG, SG, and mixed binary stars in the $r$-$x_{\rm local}(r)$ and the $r$-$E_{\rm b}$ planes for the simulation MPr5f1x3-20. This figure illustrates the interplay between segregation and binary evolution. As binaries segregate towards the inner regions they are disrupted as a result of encounters with single stars or other binaries. Harder binaries, however, can survive and evolve to become harder. SG binaries are those more affected by these evolutionary processes (ionization and hardening) as clearly shown also by  the histograms of $x_{\rm local}(r)$ and $E_{\rm b}$ in Fig. \ref{fig_xr2} and \ref{fig_ebr2}. Fig. \ref{fig_xr2} and \ref{fig_ebr2}  also show that while segregation leads to a decrease in the number of binaries in the cluster outer regions ($r>r_{\rm h}$, where $r_{\rm h}$ is the cluster half-mass radius), the distribution of binding energy of outer binaries is not evolving significantly. The effects of disruption and hardening are instead clearly visible in the distribution of binding energy of inner ( $r<r_{\rm h}$) binaries.

\begin{figure}
  \includegraphics[width=85mm]{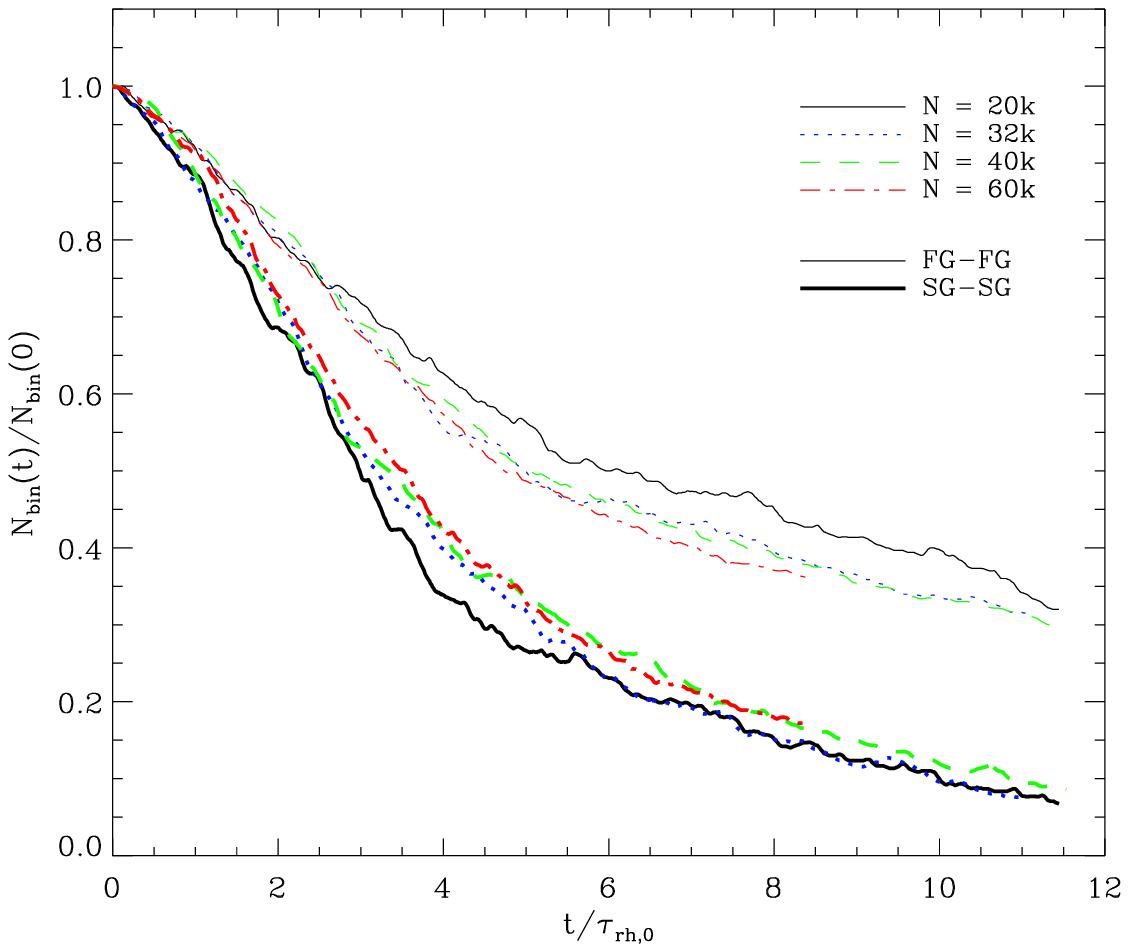}
  \caption{Time evolution of the total number of FG (thin lines) and SG (thick lines) binaries (normalized to the total initial number of binaries) for the model MPr5f03x10 and different values of the total number of particles, $N$ (see legend in the figure). Time is normalized to the initial half-mass relaxation time.
}
  \label{fig_dfn}
\end{figure}

Finally, in Fig. \ref{fig_dfn}, we show the time evolution of the total number of FG and
SG binaries for simulations with different initial number of particles (MPr5f03x10n32k, MPr5f03x10n40k, MPr5f03x10n60k). Overall the results of our simulations show a small spread
for systems with different number of particles. Since one of the
mechanisms responsible for the decrease of the number of binaries is
evaporation (although, as discussed above, it is not the dominant one)
some spread is to be expected since the dependence of the evaporation
time-scale on the number of particles differs from that of the
relaxation time (see e.g. Fukushige \& Heggie 2000, Baumgardt 2001).  

\subsection{Radial variation of the binary fraction}
In this section we focus our attention on the radial profiles of the
FG and SG binary fraction. As anticipated in the initial discussion in
\S \ref{sec:intro}, the predictions
concerning the different evolution of the global fraction of FG and SG binaries
and the general preferential disruption of SG binaries do not
necessarily imply that the local binary fraction measured at different
distances from the cluster centre follows the same trend. 

In general, the details of the evolution of the radial variation of the SG-to-FG number ratio are determined by the complex interplay among binary ionization, hardening, rate of segregation, and by differences in the initial spatial distribution of the FG and SG populations.

While, as shown in the previous section, the overall fraction of SG
binaries decreases more rapidly than that of the FG population, locally
the binary population can be dominated by SG binaries. 

\begin{figure}
  \includegraphics[width=85mm]{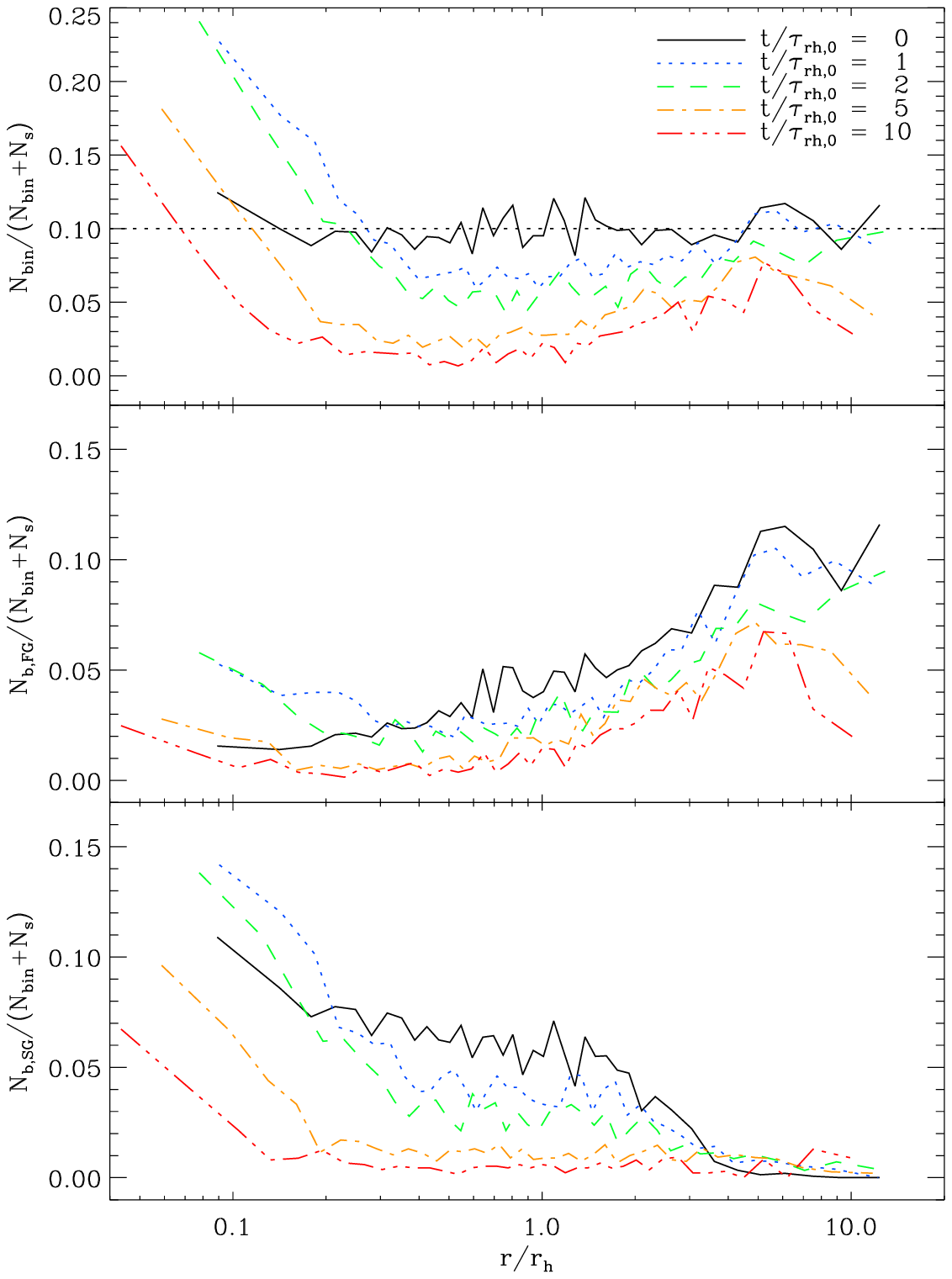}
  \caption{
Time evolution of the  radial profile of the binary fraction for the model MPr5f1x3-20 for all binaries (top panel), FG binaries  (middle panel), and SG binaries (lower panel). Different lines in each panel correspond to the times indicated in the legend in the top panel. The radial profiles have been calculated with data from three combined snapshots around the indicated time.
}
  \label{fig_bfp}
\end{figure}

\begin{figure*}
  \includegraphics[width=175mm]{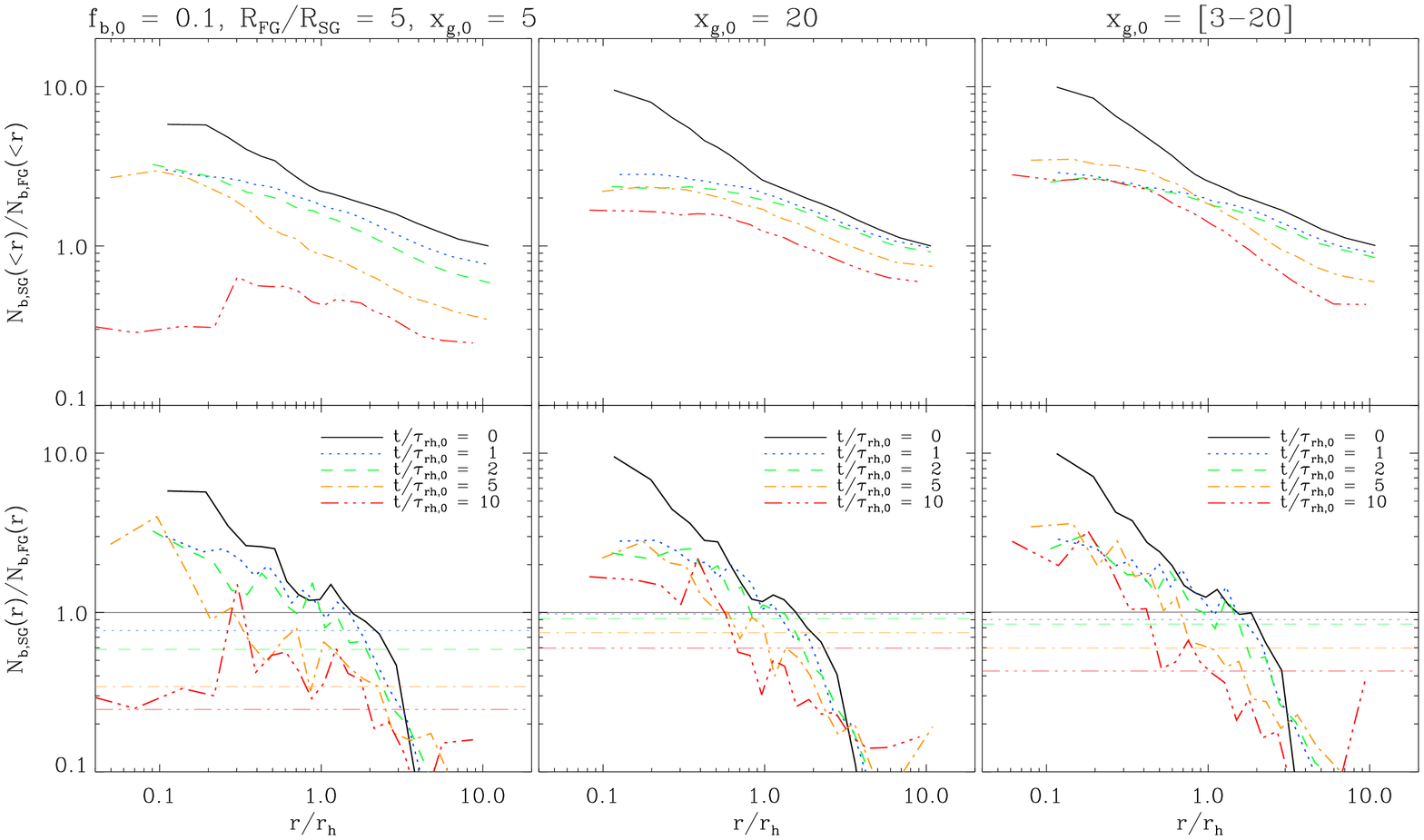}
  \caption{Top panels: time evolution of the radial profile of the ratio of the cumulative radial distributions of SG to FG binaries, $N_{\rm b,SG}(<r)/N_{\rm b,FG}(<r)$ for the simulations MPr5f1x5, MPr5f1x20, and MPr5f1x3-20. Bottom panels: time evolution of the radial profile of the SG-to-FG binary number ratio, $N_{\rm b,SG}(r)/N_{\rm b,FG}(r)$  for the simulations  MPr5f1x5, MPr5f1x20, and  MPr5f1x3-20; the horizontal lines indicate the global SG-to-FG binary number ratio . The radial profiles refer to the times shown in the legend in the lower panels. The radial profiles have been calculated with data from three combined snapshots around the indicated time (except for $t/\tau_{\rm rh,0}=10$ for which five snapshots were used).}

  \label{fig_brp}
\end{figure*}

Fig. \ref{fig_bfp}  shows the time evolution of the radial profile of the binary fraction for all the binaries (upper panel), for FG binaries (middle panel) and for SG binaries (lower panel) for the simulation MPr5f1x3-20.
This figure illustrates the effect of segregation and binary disruption: segregation towards the cluster inner regions initially leads to an enhancement of the central binary fraction, binary disruption  offsets the effects of segregation and causes the binary fraction in the central regions to decrease.

Fig. \ref{fig_brp} shows the cumulative (upper panels)
and the differential (lower panels) radial profiles of the SG-to-FG binary number ratio for the simulations MPr5f1x5, MPr5f1x20, and MPr5f1x3-20.
We point out that although the global FG binary fraction is larger
than the global SG binary fraction (see Fig. 4), in the inner regions
the binary population can still be dominated by SG binaries while
in the outer regions, the binary population is dominated by FG binaries.

Although our simulations are still focused on simplified models and not aimed at a direct comparison with observational data, it is important to emphasize the difference between the global SG and FG binary fractions and the local values of these fractions measured at different distances from the cluster centre. 
Observational studies focusing on the binary fraction in a limited range of radial distances from the cluster centre might reveal a larger fraction of FG or SG binaries depending on the radial position of the region probed by observations. The interpretation of observational results in the context of the preferential disruption of SG binaries described in this paper must therefore take into account the radial variation of the SG and FG binary fractions and the differences between local and global binary fraction of the two populations.
\begin{figure}
  \includegraphics[width=85mm]{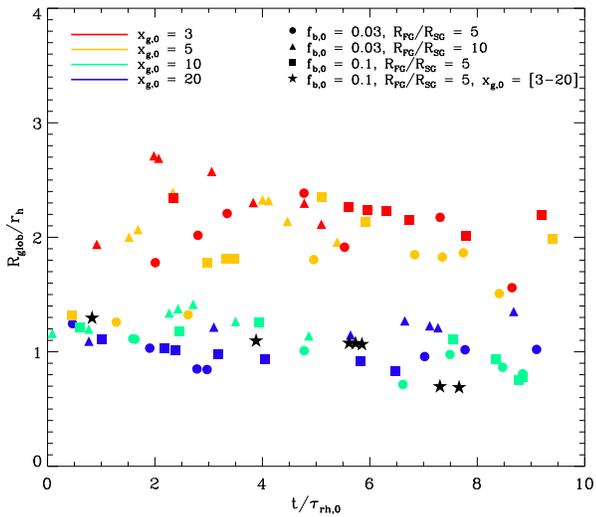}
  \caption{Time evolution of $R_{\rm glob}$, the radius where the local SG-to-FG binary number ratio equals the global value of this ratio. Different symbols and colors correspond to different simulations as indicated in the legend.}
  \label{fig_reql}
\end{figure}

While more realistic simulations are needed for a close comparison with observational studies we have further addressed this issue here and explored
the time evolution of the distance, $R_{\rm glob}$, from the cluster centre where the local FG-to-SG binary number ratio equals the global value of this ratio.
Fig. \ref{fig_reql} shows the time evolution of $R_{\rm glob}$ for all the simulations presented in this paper. $R_{\rm glob}$ does not vary significantly during the cluster evolution and it does not significantly depend on the SG initial concentration or the binary fraction. $R_{\rm glob}$  lies between approximately $r_{\rm h}$ (for simulations with  high-binding energy binaries, $\xg\geq 5$) and 2-2.5 $r_{\rm h}$ (for simulations with low-binding energy binaries which are more efficiently disrupted at larger distances from the cluster centre).
Observations focusing on this range of radial distances should therefore provide an approximate indication of the actual global value of the FG-to-SG binary number ratio. 
\section{Conclusions}
In this paper we have presented the results of a survey of $N$-body
simulations aimed at exploring the evolution of binary stars in
multiple-population globular clusters. 
In this study we have addressed a number of fundamental aspects of the dynamics of binary stars in multiple-population clusters; in subsequent studies we will further expand this investigation by including a spectrum of masses and explore how the presence of a spectrum of masses affects the interplay among mass segregation, binary ionization, hardening, and exchange interactions and the role of these processes in the dynamics of the binary population.

Theoretical models for the
formation of multiple-population clusters predict that SG stars form
segregated in the cluster inner regions and a number of observational
studies have shown that  in several clusters SG stars are more
centrally concentrated and still preserve some memory of the initial
segregation predicted by the formation models. The results of the
simulations presented  in this paper show that the initial differences
between the structural properties of the SG and the FG populations may
leave a fingerprint in the current properties of binary stars. 

We have
shown that the SG binary disruption rate is larger than that of the FG
binary population. This is a consequence of the more efficient binary
disruption   
in the cluster central high-density regions populated mainly by SG
binaries. In addition to disruption, escape also affects the number of binaries in a cluster and we have shown the relative role of these processes in driving the evolution of SG and FG binaries.

The difference between the initial spatial distributions of the SG and
the FG populations  has also implications for the properties of the
surviving binaries. Specifically, as a result of the more rapid hardening of surviving binaries in the cluster inner regions, the SG binding energy
distribution is characterized by a larger fraction of binaries with large
(more bound) binding energies. 

We have studied the evolution of the radial variation of 
the SG and FG binary fraction with the distance from the cluster centre. The
radial variation of the binary fraction 
is the result of the initial differences between the spatial
distribution of the SG and FG populations and of the subsequent effects
of binary segregation,  disruption, and escape. Although,
as the cluster evolves, SG binaries are preferentially disrupted and
the SG global binary fraction decreases more rapidly than that of the
FG binary population, the local binary fraction in the cluster inner
regions may be larger for SG binaries while the outer
regions are mostly populated by FG binaries. We have shown that the SG-to-FG
binary number ratio measured at a distance from the cluster centre equal to
about 1-2.5 $r_{\rm h}$ (where $r_{\rm h}$ is the cluster half-mass radius) is
approximately equal to the global value of this ratio. Possible
differences between SG and FG global and local binary fraction should
be  taken into account in the interpretation of observational
studies aimed at exploring the properties of SG and FG binaries spanning
limited range of distances from a cluster centre.

\section*{Acknowledgments}
EV, JH, and SLWM  acknowledge support by grants NASA-NNX13AF45G and HST-12830.01-A.
AS acknowledges the funding by the PRIN MIUR 2010-2011 'The Chemical and Dynamical Evolution 
of the Milky Way and Local Group Galaxies' (PI: F. Matteucci)
This research was supported in part by Lilly Endowment, Inc., through its support for the Indiana University Pervasive Technology Institute, and in part by the Indiana METACyt Initiative. The Indiana METACyt Initiative at IU is also supported in part by Lilly Endowment, Inc.

\end{document}